\documentstyle[pra,multicol,aps,amstex]{revtex}
\begin{document}
\setlength{\textwidth}{150mm} \setlength{\textheight}{240mm}
\setlength{\parskip}{2mm}
\input{epsf.tex}
\epsfverbosetrue
\title{Higher-order nonlinear modes and bifurcation phenomena
due to degenerate parametric four-wave mixing.}

\author{Kazimir Y. Kolossovski, Alexander V. Buryak,
Victoria V.~Steblina, and Rowland A.~Sammut}

\address{School of Mathematics and Statistics,
University College\\ Australian Defence Force Academy, Canberra
ACT 2600 Australia}

\author{Alan R. Champneys}

\address{Department of Engineering Mathematics \\ University of Bristol,
Bristol BS8 1TR, UK}

\maketitle

\begin{abstract}
We demonstrate that weak parametric interaction of a fundamental
beam with its third harmonic field in Kerr media gives rise to a
rich variety of families of non-fundamental (multi-humped)
solitary waves. Making a comprehensive comparison between
bifurcation phenomena for these families in bulk media and planar
waveguides, we discover two novel types of soliton bifurcations
and other interesting findings. The later includes (i)
multi-humped solitary waves without even or odd symmetry and (ii)
multi-humped solitary waves with large separation between their
humps which, however, may not be viewed as bound states of several
distinct one-humped solitons.
\end{abstract}


\begin{multicols}{2}
\narrowtext
\section{Introduction and Model}
Recently parametric wave mixing in Kerr media has attracted
significant attention (see, e.g., Refs. \cite{saltiel,ours,ours2D}
where continuous wave (CW) interaction and parametric
self-trapping were investigated). This theoretical activity has
been backed up by experimental advances, e.g., a novel scheme for
quasi-phase matched third harmonic generation (THG) has been
suggested \cite{will}. However, in previous works devoted to
spatial solitary waves due to THG in planar waveguides, only
families of fundamental self-trapped beams were considered.  In
the Letter \cite{ours2D}, for example, where solitons due to third
harmonic generation were considered for a bulk medium geometry,
higher-order modes were not discussed in detail. By {\em
higher-order} we refer to beam shapes whose transverse intensity
typically has a multi-peaked structure and has higher energy than
the single-peaked fundamental state.

In this work we analyze in some detail the structure and
bifurcation phenomena of higher-order bright spatially localized
modes or `solitons', which we do not use in a strict mathematical
sense, since the models in question are not integrable. The
spatial configuration is assumed to be such that there is a well
defined propagation direction and the beams are localized in $n$
transverse directions, with $n=1$ representing a planar waveguide
and $n=2$ a bulk medium. Specifically we study models representing
(1+1)-dimensional and (2+1)-dimensional, weakly-anisotropic media
with cubic nonlinearity, under the phase-matched condition that
the fundamental wave is resonantly coupled to its third harmonic.
This is a particular degenerate case of solitons supported by the
four-wave mixing processes \cite{fwm}, which is not completely
understood yet in full generality.  We assume that the interaction
between the fundamental and third-harmonic  waves includes the
effects of parametric four-wave mixing, self-phase modulation, and
cross-phase modulation.

We closely follow the derivation procedure of Ref.~\cite{ours},
assuming that the fundamental and the third-harmonic beams have
the same linear polarization. The result is the following
normalized (dimensionless) system of coupled equations:

\begin{equation}
\label{eq_u}
\begin{array}{l}
{\displaystyle i \frac{\partial u}{\partial z} +\nabla^2 u - u +
\left( \frac{1}{9} |u|^2 + 2 |w|^2 \right) u + \frac{1}{3} u^{*2}
w = 0,} \\ \\ {\displaystyle i \sigma \frac{\partial w}{\partial
z} +\nabla^2 w
 - \alpha w + (9 |w|^2 + 2 |u|^2) w + \frac{1}{9} u^3 = 0,}
\end{array}
\end{equation}
where $u$ and $w$ are the fundamental and third harmonics,
respectively. Also for the case of spatial beams $ \nabla^2 \equiv
\partial^2/ \partial x^2 + \partial^2/
\partial y^2$ in the (2+1)-dimensional case, or $ \nabla^2 \equiv
\partial^2/ \partial x^2$ in the (1+1)D case. The parameter
$\alpha$ measures the shift in the propagation constant, which is
induced by the nonlinearity and is also dependent on the quality
of wave-vector matching between the harmonics, with $\alpha = 3
\sigma$ corresponding to exact matching, and  $z$ is the
propagation distance. For the spatial soliton case the
dimensionless parameter $\sigma$ is the ratio of the wave numbers
of the harmonics and is equal to $3$. Note that the system
(\ref{eq_u}) may also describe temporal pulse propagation of
resonantly interacting fundamental and third harmonics in optical
fibers. For this physical situation $\nabla^2 \equiv \partial^2/
\partial t^2$ ($t$ is the retarded time variable) and $\sigma$ is
the absolute value of the ratio of second-order group velocity
dispersions for the first and the third harmonics and may be any
positive number.

Radially symmetric stationary beams are described by real
solutions $u(r)$ and $w(r)$ which are defined by the following
system:

\begin{equation}
\label{eq_st}
\begin{array}{l}
{\displaystyle  \frac{d^2 u}{d r^2} + \frac{s}{r} \frac{d u}{d r}-
u + \left( \frac{1}{9} u^2 + 2 w^2 \right) u + \frac{1}{3} u^2 w =
0,} \\*[9pt] {\displaystyle \frac{d^2 w}{d r^2} + \frac{s}{r}
\frac{d w}{d r}
 - \alpha w + (9 w^2 + 2 u^2) w + \frac{1}{9} u^3 = 0.}
\end{array}
\end{equation}

Here $r \equiv \sqrt{x^2+y^2}$ and $s = 1$ for the (2+1)D case,
whereas $r  \equiv x$ and $s= 0$ for the (1+1)D case. These
localized solutions depend only on a single dimensionless
parameter $\alpha$. Analysis of the linear part of Eqs.
(\ref{eq_st}) in the limit $r \to \infty$ shows that conventional
bright solitons (with exponentially decaying tails) can exist only
for $\alpha > 0$.

By `bright symmetric' in the remainder of this paper we shall mean
(2+1)D solutions when the intensity of each localized harmonic
reaches a maximum at $r=0$ and (1+1)D solutions with $u(r) =
u(-r)$ and $w(r) = w(-r)$. Thus, we shall only seek these
solutions on the interval $0\leq r \leq \infty$ even in the
$(1+1)D$ case. Note further that Eqs.\ (\ref{eq_st}) have odd
symmetry, that is, if $[u(r),w(r)]$ is a solution then so is
$[-u(r),-w(r)]$. Thus all solutions must come in pairs, the second
solution being simply a change in sign (a phase shift of $\pi$) of
both harmonics. For the case $s=0$, it is additionally possible to
have solutions which are odd in both harmonics, or which are
neither odd nor even. The latter type of solutions we shall refer
to as being `bright asymmetric'. In this paper we shall consider
mainly the solitons of bright symmetric type, but shall also
present some results about bright asymmetric (1+1)D solitons. Dark
solitons (localized solutions with nonzero asymptotics) are out of
the scope of this paper.

The case $\alpha<0$ also has physical meaning, but there one
should expect to find {\em quasisolitons}, which are almost
localized stationary states that have small periodic oscillations
in their tails. See e.g. \cite{ours2D,quas_st,radiation,lombardi}
for the definition, examples and for some issues surrounding them.
Quasisolitons in this model will form the subject of another work.
Here we shall concentrate almost exclusively on the case
$\alpha>0$.

Using a direct analogy with the theory of $\chi^{(2)}$
(second-harmonic generation or SHG) solitons (e.g.
Ref.~\cite{SHG}), we start our analysis from the so-called
cascading limit when $\alpha \gg 1$. In this limit $w \approx
u^3/(9 \alpha)$ and the equation for $u$ approaches the
cubic-quintic Nonlinear Schr\"{o}dinger (NLS) equation. This
scalar equation possesses a familiar class of fundamental bright
solitons consisting of a simple bell-shape [there are also
higher-order families in the (2+1)D case]. These fundamental
solitons can then be used as a starting point in the search for
families of stationary solutions using numerical methods. These
methods comprise a standard shooting method at fixed $\alpha$, and
a continuation method allied to solution using a relaxation method
for solving an appropriately defined two-point boundary-value
problem for Eqs.~(\ref{eq_st}). This latter technique can trace
paths of solutions as $\alpha$ varies. We choose to characterize
these solitons by the value of normalized total power which is one
of the conserved quantities of the system (\ref{eq_u})
\begin{equation}
\label{power} P_{\rm tot} = \int_{A}\, (|u|^2 + 3 \sigma |w|^2) \,
d A.
\end{equation}
Here the integration extends over the appropriate one or
two-dimensional infinite cross-section $A$.

The dependence of $P_{\rm tot}$ on $\alpha$ for a branch of
solitons is usually, at least in the case of a fundamental
solution, closely related to its stability. A necessary condition
for stability in the case of fundamental multi-component solitons
is typically given by a generalized Vakhitov-Kolokolov (VK)
criterion \cite{vahitov}, which often also appears to be  a
sufficient condition for soliton stability (see, e.g.,
Refs.~\cite{VK_examp}). However, the complexity of
Eqs.~(\ref{eq_u}) which, for example, possess collapse-type
dynamics in the (2+1)D case, may lead to instability of
fundamental solitons even for branches which are supposed to be
stable according to the VK criterion~\cite{ours2D}. Thus, below we
use power versus-$\alpha$ dependence only for classification of
soliton families, leaving a full-scale stability analysis for
future consideration.

\section{Results for bulk media}
\begin{figure}[h]
\setlength{\epsfxsize}{7.5cm} \centerline{\epsfbox{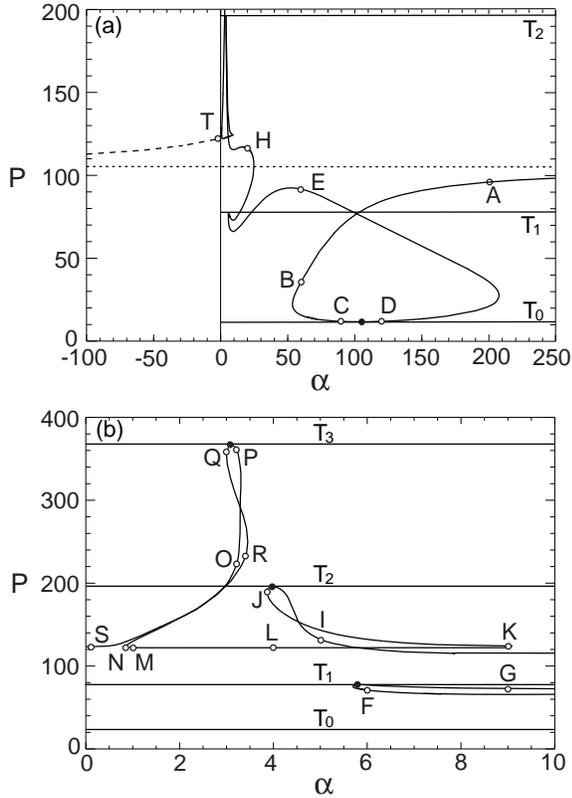}}
\caption{(a) Bifurcation diagram for solitons (solid curves) and
quasisolitons (dashed curves) of Eqs.~({\protect{\ref{eq_u}}}) in
the (2+1)D case. (b) Expanded portion of Fig.~\ref{main1}(a) for
the range $0\leq\alpha\leq10$. Examples of solitons are shown in
Figs.~\ref{fig2}--\ref{fig7}. Bifurcation points of two-wave
solitons from one-wave soliton families are shown by filled
circles. The results related to quasisolitons are for stationary
solutions with {\em minimal} amplitude of oscillatory tails; in
that case, $P$ is calculated for the soliton core only.}
\label{main1}
\end{figure}

\begin{figure}[h]
\vspace*{-8mm} \setlength{\epsfxsize}{7.5cm}
\centerline{\epsfbox{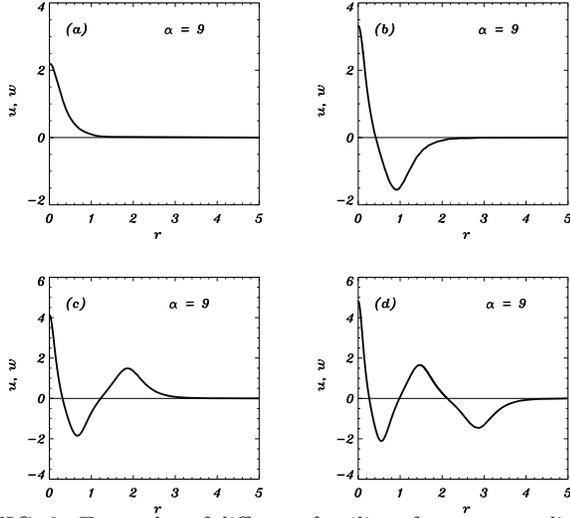}} \caption{Examples of different
families of one-wave solitons. In all diagrams, the thick line
corresponds to the $3^{rd}$ harmonic.} \label{fig2}
\end{figure}

\begin{figure}[h]
\setlength{\epsfxsize}{7.5cm} \centerline{\epsfbox{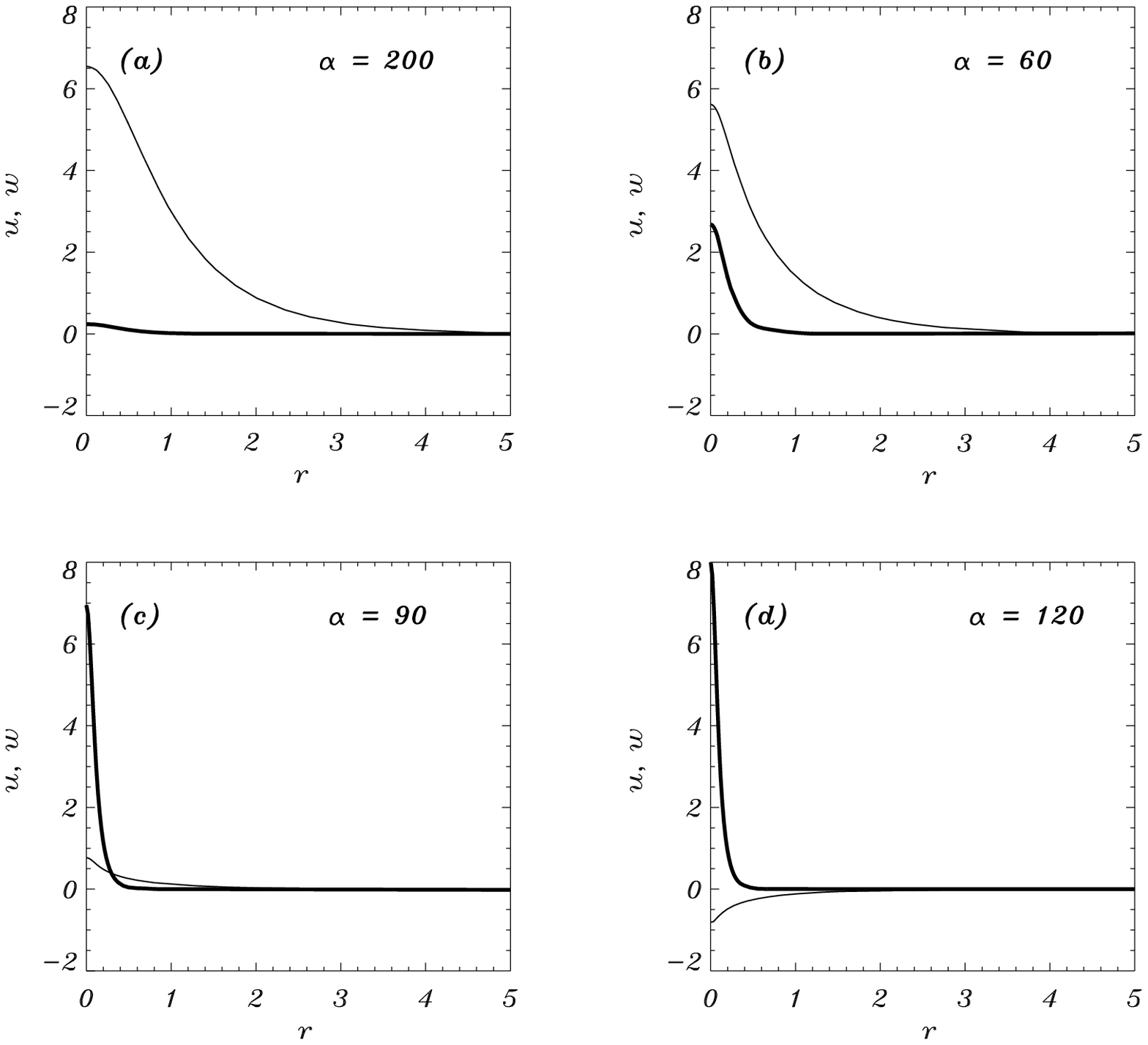}}
\caption{Examples of (2+1)D two-wave solitons. Labeling of
examples corresponds to labeling of open circles in
Fig.~\ref{main1}} \label{fig3}
\end{figure}

\begin{figure}[h] \setlength{\epsfxsize}{7.5cm}
\centerline{\epsfbox{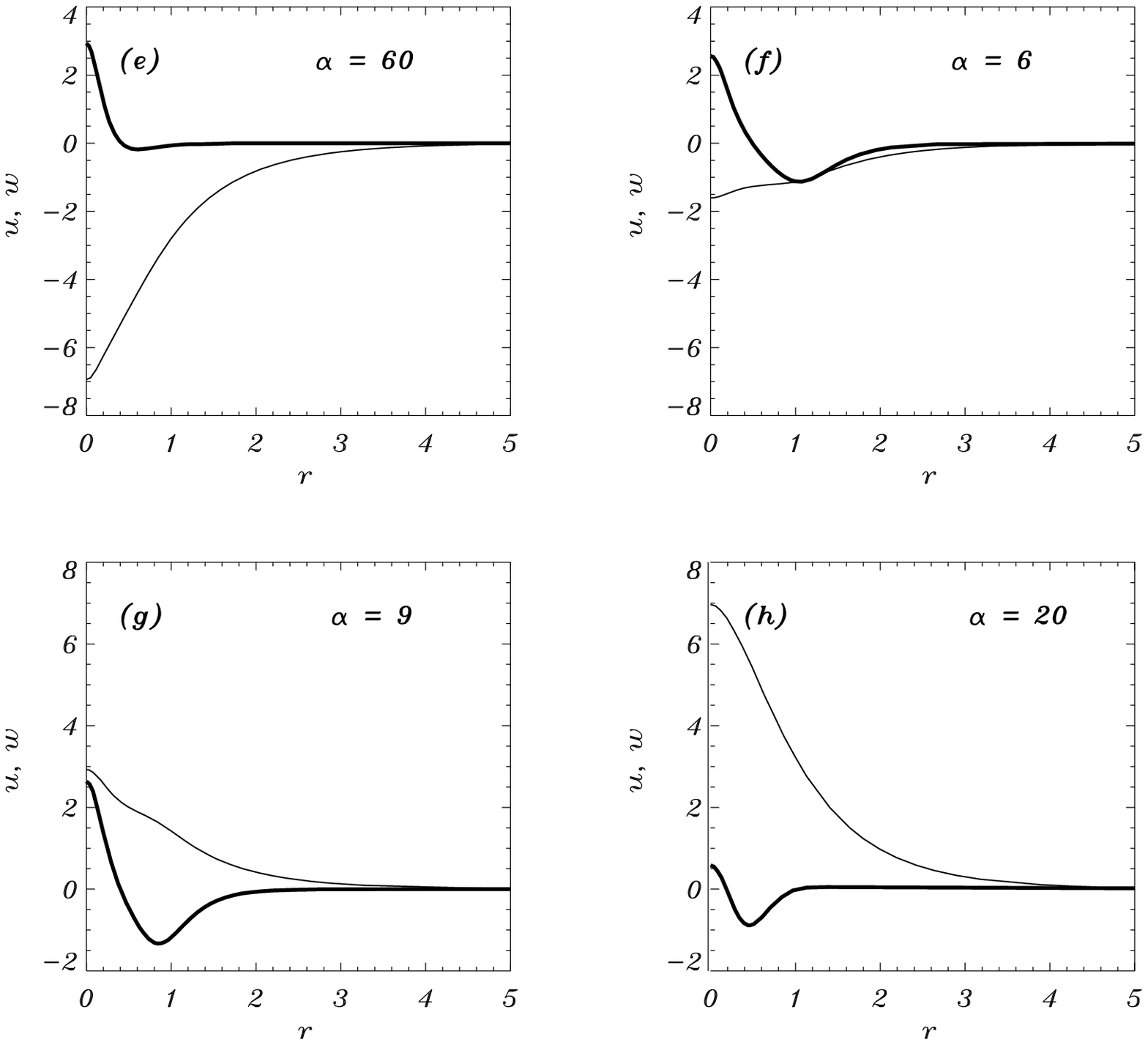}} \caption{Examples of (2+1)D
two-wave solitons. Labeling of examples corresponds to labeling of
open circles in Fig.~\ref{main1}} \label{fig4}
\end{figure}

\begin{figure}[h] \setlength{\epsfxsize}{7.5cm}
\centerline{\epsfbox{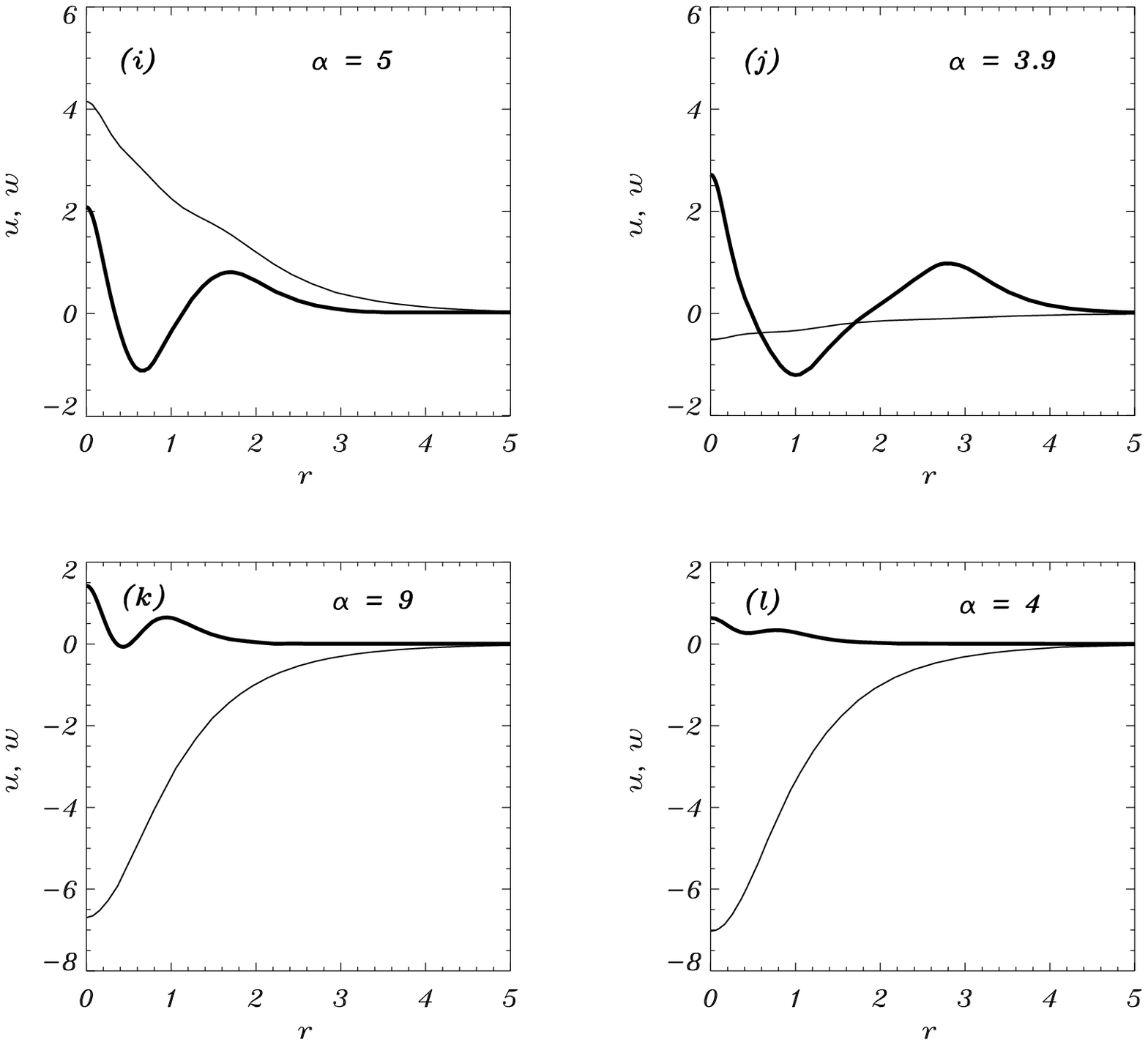}} \caption{Examples of  (2+1)D
two-wave solitons.  Labeling of examples corresponds to labeling
of open circles in Fig.~\ref{main1}} \label{fig5}
\end{figure}

\begin{figure}[h]
\setlength{\epsfxsize}{7.5cm} \centerline{\epsfbox{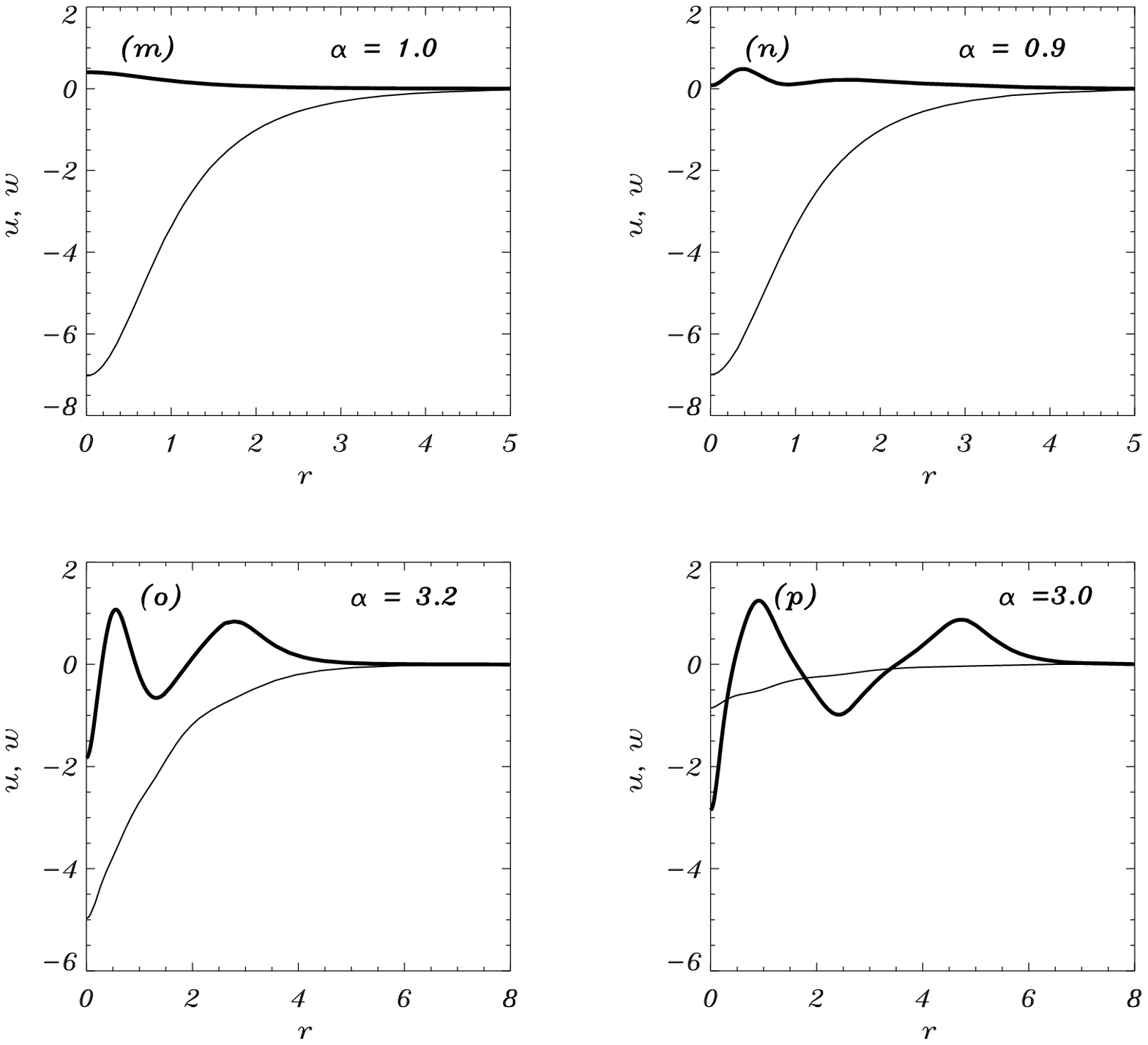}}
\caption{Examples of  (2+1)D two-wave solitons. Labeling of
examples corresponds to labeling of open circles in
Fig.~\ref{main1}} \label{fig6}
\end{figure}

\begin{figure}[h]
\setlength{\epsfxsize}{7.5cm} \centerline{\epsfbox{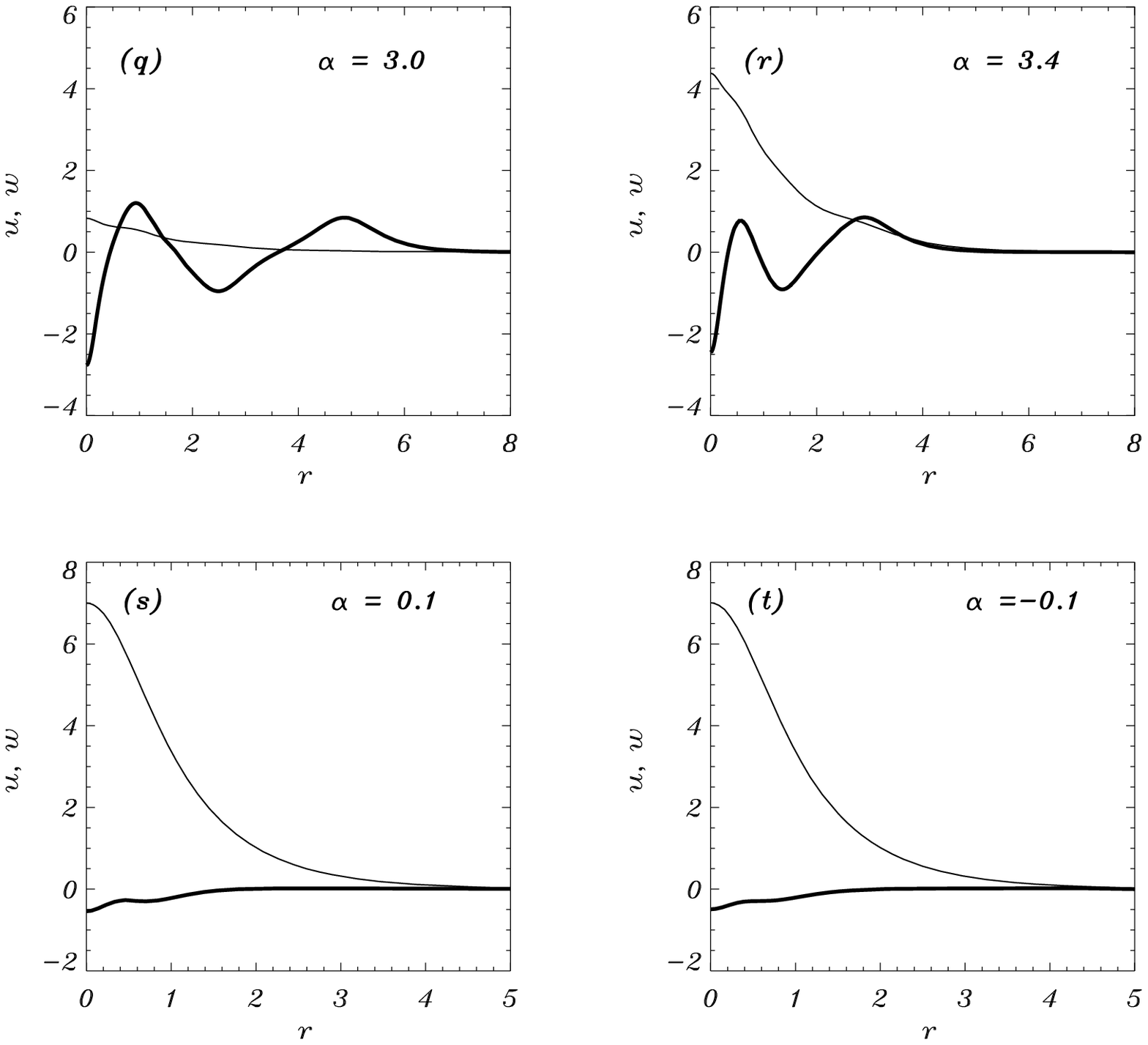}}
\caption{Examples of  (2+1)D two-wave solitons and quasisolitons.
Labeling of examples corresponds to labeling of open circles in
Fig.~\ref{main1}}\label{fig7}
\end{figure}

First we present the results for the (2+1)D case.
Figure~\ref{main1} shows the variation of the normalized total
power, $P_{\rm tot}$, with the normalized mismatch parameter
$\alpha$, for different types of one-wave and two-wave localized
solutions of the system~(\ref{eq_u}) with $s=1$. The corresponding
soliton profiles at various points along the presented branches
are given in Figs.~\ref{fig2}--\ref{fig7}.

The first class of localized solutions of the system (\ref{eq_u})
consists of one-frequency soliton families for the third harmonic
$w_0$, which exist for all $\alpha >0$ and represent scalar Kerr
solitons described by the standard cubic NLS equation which
follows from the second of Eqs.~(\ref{eq_u}) at $u=0$:
\begin{equation}
\label{ti}
\begin{array}{l}
{\displaystyle \frac{d^2 w_0}{d r^2} + \frac{1}{r} \frac{d w_0}{d
r}-\alpha w_0 + 9 w^3_0 = 0.}
\end{array}
\end{equation}
These single frequency solitons differ from each other by the
number of zero crossings in their radial profiles so that we
denote the corresponding families as $T_0$ (no crossing), $T_1$
(one crossing), $T_2$ (two crossings), etc. Examples of one-wave
solitons belonging to different $T_j$ families are shown in
Fig.~\ref{fig2}. Note that the normalized power $P_{\rm tot}$ is
constant for each of the $T_i$ families. For example, for the
fundamental one-wave soliton family $T_0$ (which are, in fact,
Townes solitons of Ref. \cite{townes}) we have $P_{tot} \approx
11.70$ for all $\alpha>0$.

The second class of solutions to Eqs.~(\ref{eq_u}) are genuinely
two-wave bright solitons, described by families of localized beams
with coupled fundamental and third harmonics. The simplest way to
obtain such solutions numerically is to follow the two-wave
soliton families from the cascading limit (large $\alpha$) as
$\alpha$ decreases.  In this work we concentrate on the result of
following the lowest order two-wave soliton branch whose profiles
have a simple one-hump form in the cascading limit. For this
family, painstaking numerical continuation reveals a highly
complex solution path involving restructuring of the soliton
profile while the corresponding $P(\alpha)$ curve undergoes
several loops (see Figs.~\ref{main1} and  \ref{fig1-1}). Inherent
in each loop is a touch with one of the $T_i$ families. Such a
touch corresponds to a transcritical bifurcation from the pure
$w$-solution, and note [for example from Fig.~\ref{fig3}(c),(d)
which correspond to points C and D on Fig.~\ref{main1}(a)] that
the two different bifurcating branches have opposite signs of
their $u$-component. The fact that these bifurcations take place
further illustrates the severity of the restructuring of the
soliton profiles that must take place; in the cascading limit the
branch is approximately of pure $u$-type, whereas at each
bifurcation with $T_i$ it is composed of purely a third harmonic
component $w$. Figures~\ref{fig2}--\ref{fig7} illustrate the
complete restructuring process by depicting the soliton profiles
in the vicinity of each bifurcation and turning point of the
$P(\alpha)$ curve. Note finally that the two-wave soliton family
also includes the simplest so-called self-similar [for which $u(r)
\propto w(r)$] solution (Fig.~\ref{main1}, point M) at $\alpha=1$,
see Ref.~\cite{koshiba} for the details and also Ref.~\cite{ours},
where its (1+1)-dimensional counterpart was also been considered.

The position of the bifurcation point from the $T_0$ branch can be
approximately calculated analytically. Linearization of
Eqs.~(\ref{eq_u}) around the solution $w_0(r)$ gives the
eigenvalue equation
\begin{equation}
\label{eq_lin1}
\begin{array}{l}
{\displaystyle \frac{d^2 u_1}{d r^2} +  \frac{1}{r} \frac{d u_1}{d
r} + 2w_0^2(r) u_1 = \lambda u_1},
\end{array}
\end{equation}
together with appropriate boundary conditions. Bifurcations occur
when $\lambda=1$. This may also be viewed as the problem of
existence of localized states in the potential $U(r)=-2w_0^2(r)$
with eigenvalue $\lambda$. Due to the lack of a closed form
analytical expression for $w_0^2(r)$, solutions of (\ref{eq_lin1})
may be approximated by feeding in the numerical data for $w_0$ or
by analytical techniques based on a variational approximation.
Using the latter, based on a simple exponential trial function,
gives the result $w_0=\sqrt{8\alpha}/3e^{-r\sqrt{\alpha}}$.
Substituting this into Eq.~(\ref{eq_lin1}) and assuming a similar
form of trial function for $u_1$, one can use a Rayleigh-Ritz
method to obtain $\alpha^{(var)}_{bif} = 105.8$.  This agrees to
within 2\% with the numerical result $\alpha^{(num)}_{bif} =
104.2$. Calculation of bifurcation points along the higher-order
$T_i$ branches may in principle be carried out by the same method.
However, this is less straightforward technically because it
requires the use of complicated forms of trial functions, and it
is perhaps more instructive to rely on numerical detection of the
bifurcation points. Symmetry arguments dictate that at each
bifurcation point $\alpha_{bif}$ there will exist two different
bifurcating solutions of Eq.~(\ref{eq_lin1}): $u_1(r)$ and
$-u_1(r)$. Moreover, each bifurcation is a transcritical and gives
rise to a pair of two-wave branches $(w_0(r),\pm \varepsilon
u_1(r))$, where $\varepsilon$ is proportional to
$|\alpha-\alpha_{bif}|$.

The third class of solutions to Eqs.~(\ref{eq_u}) are the
aforementioned quasisolitons which exist in the region of negative
$\alpha$. We do not discuss quasi-solitons here in any detail. A
full analysis will appear elsewhere. We simply make the comment
that the branch SOP bifurcating from $T_3$ can be continued up to
the boundary $\alpha=0$ separating regular from quasi-solitons. On
the other side of the boundary a similar quasi-soliton state can
be found with tiny oscillations in its tail [see Fig.~\ref{main1}
and Fig.~\ref{fig7}(t,s)].

\newpage
\vspace*{-3mm}
\vspace*{-3mm}
\vspace*{-12mm}
\vspace*{-3mm}

\newpage
\section{Results for planar waveguides}
\begin{figure}[h]
\setlength{\epsfxsize}{7.5cm} \centerline{\epsfbox{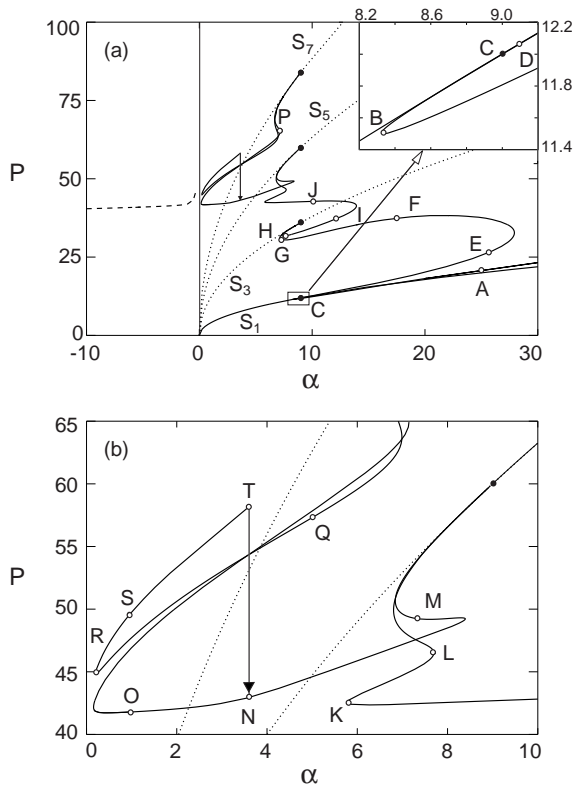}}
\caption{(a) Bifurcation diagram for symmetric solitons (solid
curves) and quasisolitons (dashed curve) of the (1+1)D version of
Eqs.~({\protect{\ref{eq_u}}}). (b) Expanded portion of
Fig.~\ref{main2}(a) in the range $0\leq\alpha\leq10$, $40\leq
P\leq65$. Dotted curves emerging at zero correspond to integer
multiples of the primary one-wave solution $S_1$. Formally they
represent multi-soliton states consisting of a concatenation of
infinitely separated single solitons. Points at which branches of
two-wave solitons terminate by `bifurcating' from one of these
multi-solitons are depicted by filled circles and all occur for
$\alpha=9$. The inset to (a) and the jump ($T \to N$) depicted in
(b) are explained in the text.} \label{main2}
\end{figure}

\begin{figure}[h]
\setlength{\epsfxsize}{7.5cm} \centerline{\epsfbox{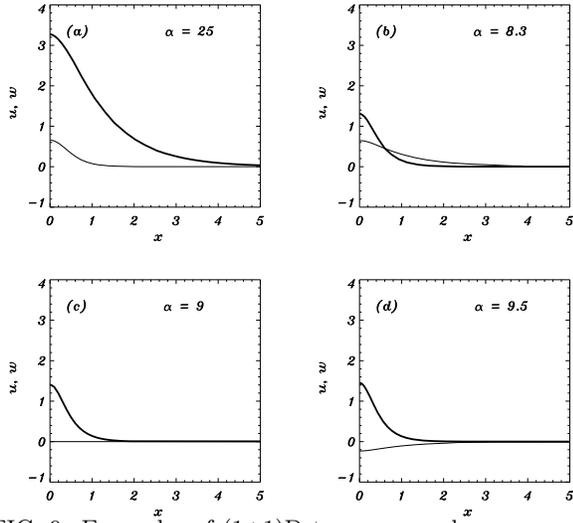}}
\caption{Examples of (1+1)D two-wave and one-wave solitons.
Labeling of all examples corresponds to the labeling of the open
circles in Fig.~\ref{main2}.} \label{fig9}
\end{figure}

\begin{figure}[h]
\setlength{\epsfxsize}{7.5cm} \centerline{\epsfbox{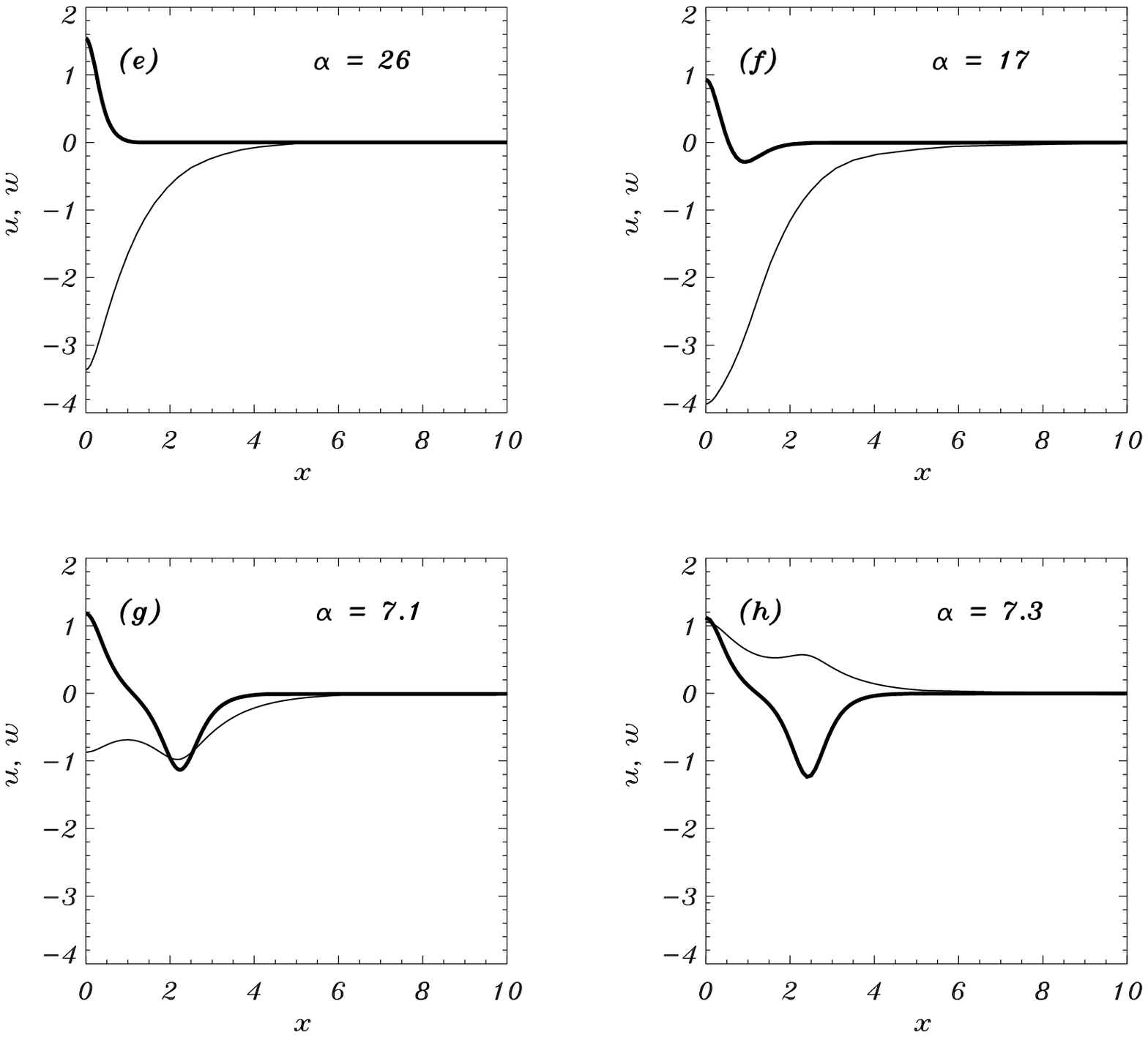}}
\caption{Examples of (1+1)D two-wave solitons. Labeling is as for
Fig.~\ref{fig9}.} \label{fig10}
\end{figure}

\begin{figure}[h]
\setlength{\epsfxsize}{7.5cm} \centerline{\epsfbox{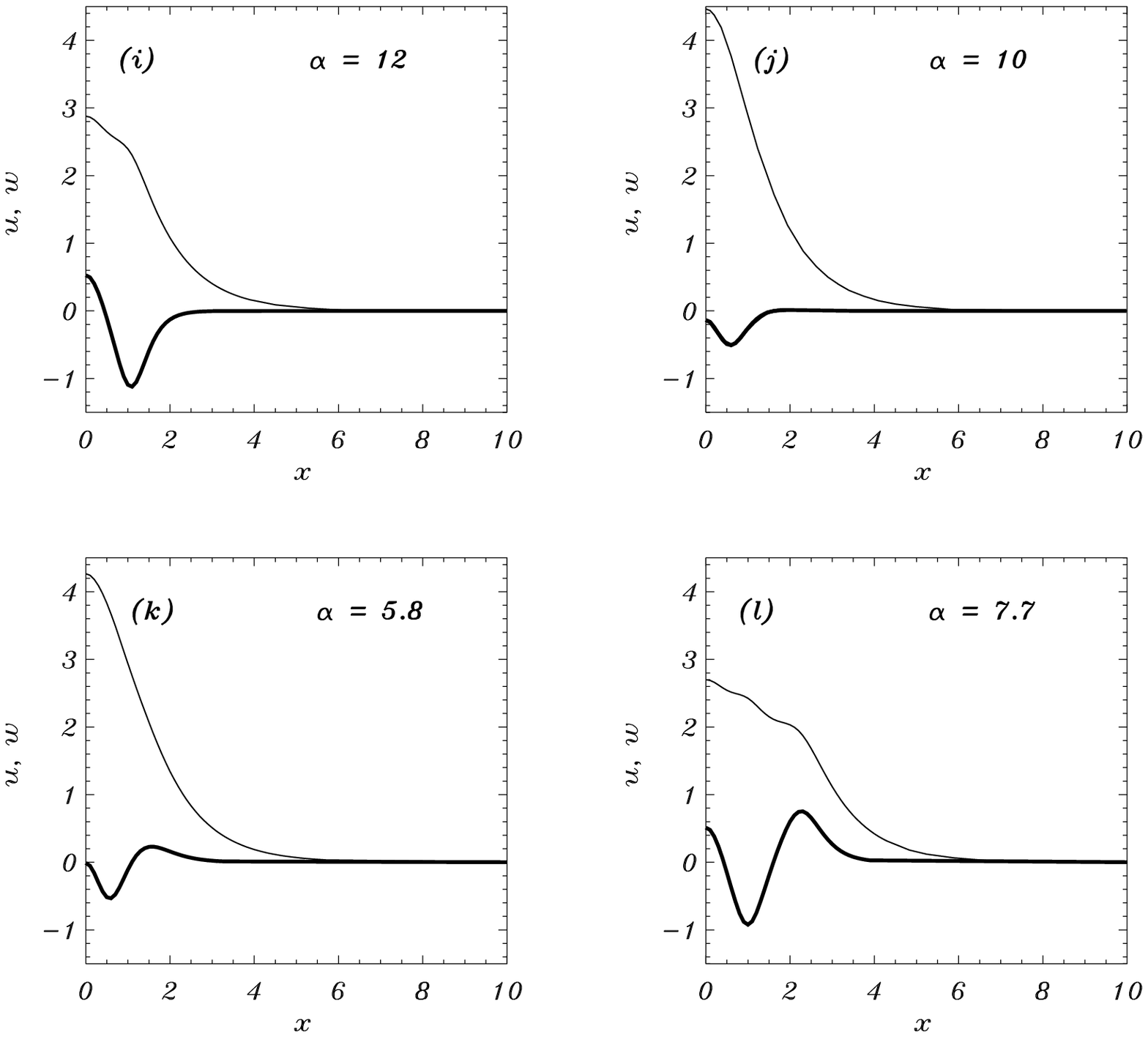}}
\caption{Examples of (1+1)D two-wave solitons. Labeling is as for
Fig.~\ref{fig9}.} \label{fig11}
\end{figure}

\begin{figure}[h]
\setlength{\epsfxsize}{7.5cm} \centerline{\epsfbox{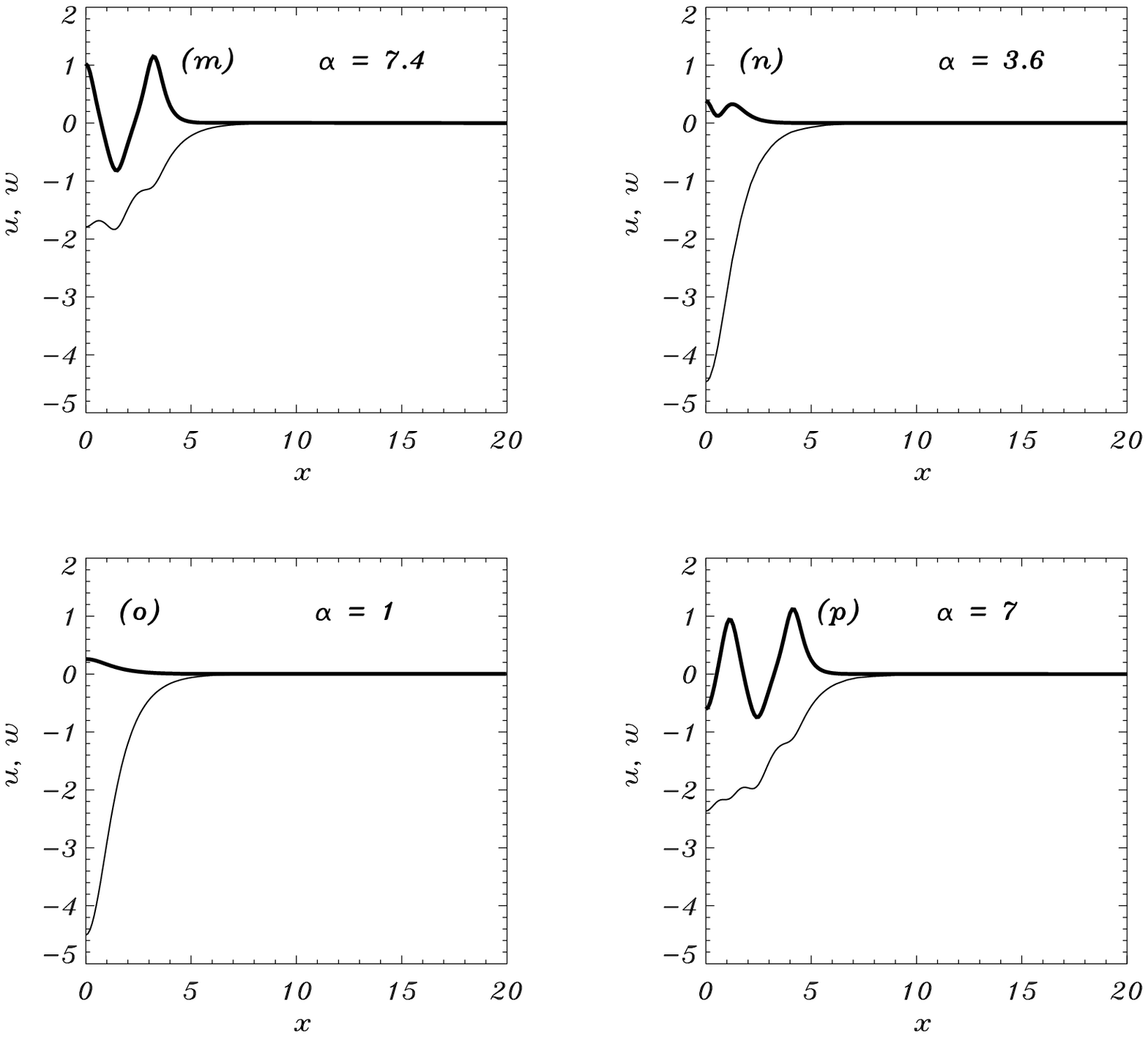}}
\caption{Examples of (1+1)D two-wave solitons. Labeling is as for
Fig.~\ref{fig9}.} \label{fig12}
\end{figure}

\begin{figure}[h]
\setlength{\epsfxsize}{7.5cm} \centerline{\epsfbox{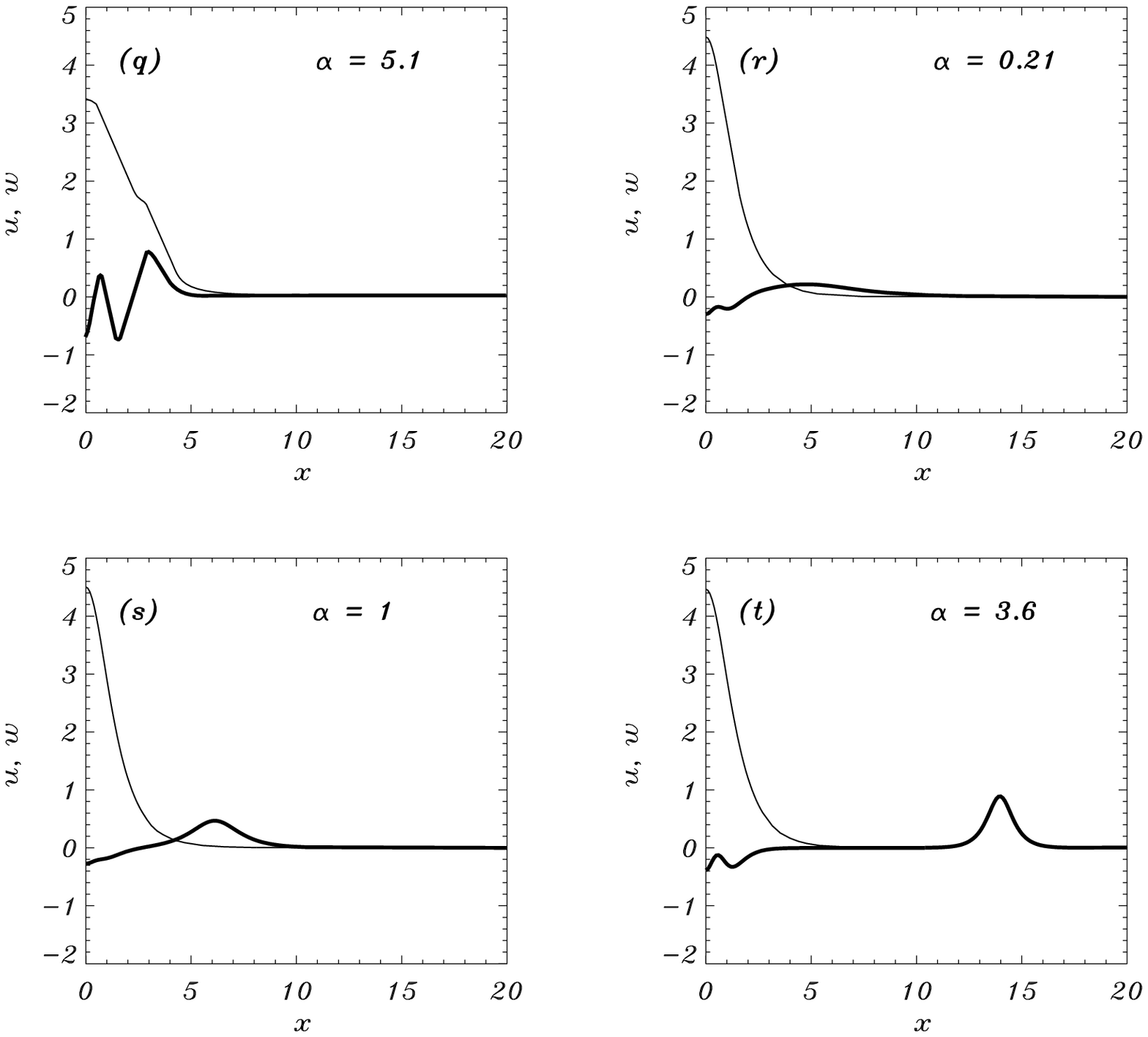}}
\caption{Examples of (1+1)D two-wave solitons. Labeling is as for
Fig.~\ref{fig9}.} \label{fig13}
\end{figure}

\begin{figure}[h]
\setlength{\epsfxsize}{7.5cm} \centerline{\epsfbox{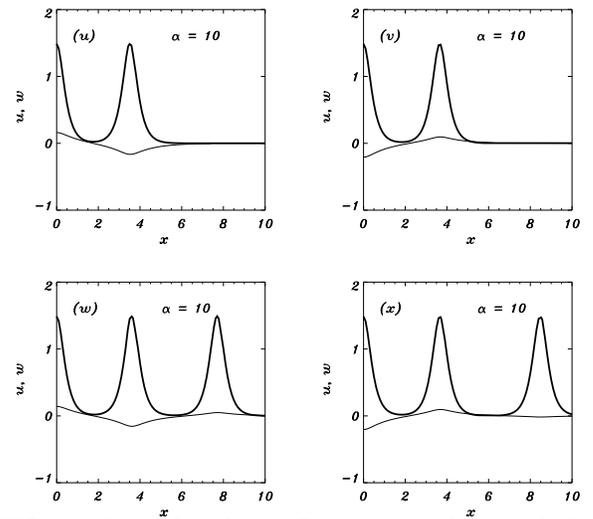}}
\caption{Examples of (1+1)D two-wave solitons, which are not
directly linked to the two-wave solitons of the cascading limit.
Labeling is as for Fig.~\ref{fig9}.} \label{fig14}
\end{figure}

It is interesting to compare the (2+1)D results discussed above
with those for the corresponding (1+1)D case. The bifurcation
diagram related to the (1+1)D case is presented in
Fig.~\ref{main2} and the corresponding examples of soliton
profiles are given in Fig.~\ref{fig9}--\ref{fig14}. We now
highlight how, together with many obvious differences in
comparison to the diagram for the (2+1)D case in Fig.~\ref{main1},
there are also some striking similarities as well. Note that in
some respects the model for the (1+1)D case is simpler since the
corresponding stationary system (\ref{eq_st}) with $s=0$ does not
depend explicitly on $r$ and hence represents an autonomous
dynamical system in four dimensions. Finding solitons is then
reduced to finding homoclinic trajectories in this 4D phase space.

The first class of (1+1)D localized waves of system (\ref{eq_u})
consists of one-frequency soliton families for the third harmonic
$w_0$, which exist for all $\alpha >0$ and represent scalar Kerr
solitons described by the standard cubic (1+1)D NLS equation which
follows from the second of Eqs. (\ref{eq_u}) at $u=0$:
\begin{equation}
\label{tii}
\begin{array}{l}
{\displaystyle \frac{d^2 w_0}{d x^2}-\alpha w_0 + 9 w^3_0 = 0.}
\end{array}
\end{equation}
It can be readily solved exactly giving the well-known unique
single soliton solution:
\begin{equation}
\label{wo} w_0(x)=\frac{\sqrt{2\alpha}}{3} \mbox{sech}
(\sqrt{\alpha}x), \;\;\;\; P_{tot}=4\sqrt{\alpha}.
\end{equation}

In contrast to the (2+1)D case strictly speaking there are no
other one-wave localized solutions. However, it will be helpful in
what follows to consider formal multi-soliton states consisting of
a different number of infinitely separated single solitons
(\ref{wo}), families of which we denote by $S_1$ (single soliton),
$S_2$ (two solitons), $S_3$ (three solitons), etc. In this work we
are mainly interested in families with an odd number of separated
solitons: $S_{2i+1}, \; i=1,2,3,\ldots$, but we also investigate
`bifurcations' from $S_2$. Note that, for $i > 1$, $S_i$ in fact
denotes more than a single one-wave family, because each single
pulse that is glued together can be either positive or negative.

The second class of (1+1)-dimensional localized solutions of
Eqs.~(\ref{eq_u}) consists of two-wave bright symmetric solitons
and is described by families of localized beams with coupled
fundamental and third harmonics. The simplest way to obtain  the
lowest order two-wave soliton family is again to continue
numerically from solitons of the cascading limit ($\alpha \gg 1$)
given approximately by the expression:
\begin{equation}
\label{casc}
\begin{array}{l}
{\displaystyle u(x) \approx  \frac{6}{\sqrt{1+B\; {\rm
cosh}\;2x\;\;}}},\;\; {\displaystyle w \approx {u^3}/(9\alpha),}
\end{array}
\end{equation}
where $B=\sqrt{1+16/\alpha}$. The first expression for $u(x)$ in
Eq.~(\ref{casc}) is the solution of the corresponding
cubic-quintic NLS-type equation.

The results of our numerical continuation from this limiting
solution, upon decreasing $\alpha$ is that, like in the (2+1)D
case, this branch also traces a convoluted path in the
$(P,\alpha)$-plane, involving four `bifurcations' from one-wave
soliton families (from the families $S_1$, $S_3$, $S_5$, and
$S_7$). As in the (2+1)D case, this branch connects to a
self-similar solution at $\alpha=1$ (the point O in
Fig.~\ref{main2}(b)). In this case, the self-similar solution is
expressible in closed analytical form as
\begin{equation}
\label{selfsim}
 u(x) = a \;{\rm sech} \; x, \;\;\;\; w(x) = b\; u(x),
\end{equation}
where the parameter $b$ is the real root of the cubic equation $63
b^3 - 3 b^2 + 17 b + 1 = 0$, and $a^2 = 18/(18 b^2 + 3 b + 1)$.
However, it is here that the similarity with the $(2+1)$-case
ends, as we shall now explain.

\newpage
\vspace*{-14mm}
\vspace*{-3mm}
\vspace*{-3mm}
\vspace*{-14mm}
\vspace*{2mm}
\vspace*{-3mm}

First, let us try to motivate what is happening at each of the
`bifurcations' from $S_j$; for which at first sight it seems
remarkable that each one occurs precisely at $\alpha=9$. Standard
bifurcation analysis (e.g.\ as in Ref. \cite{nail}) allows us to
find the position of the single bifurcation point from the
one-wave soliton family $S_1$ (\ref{wo}) at $\alpha = 9.0$ (point
C in Fig.~\ref{main2}(a)). As in the (2+1)D case the bifurcation
is a transcritical with one branch emerging to the left of the
bifurcation point and one to the right. This structure is
confirmed by the inset to Fig.~\ref{main2}(a) which shows that the
branch emerging to the left undergoes a fold (at point B), so that
on a larger scale both branches appear to bifurcate to the right.

Now it seems that this `local' bifurcation from $S_1$ causes a
topological change in the four-dimensional phase space so that a
global event must also happen at this parameter value. This global
event is the possibility of gluing together several copies of the
$S_1$ back to back and forming a new branch of solitons with
several large peaks that bifurcate from $\alpha=9$.
Phenomenologically this is similar to what happens in the SHG case
when the parameter equivalent to $\alpha$ passes through 1
\cite{yew+,yew}. A key observation here is that in order to get a
symmetric (even) solution, only an odd number of copies of the
$S_1$ may be taken to form solitons in this way. As a convenient
short-hand for this global bifurcation of multi-peaked solutions
at $\alpha=9$, we have refereed to it as a local `bifurcation'
from $S_{2i+1}$ where $i=1,2,3\ldots$, although this is strictly a
misnomer.

Numerical continuation beyond point G of Fig.~\ref{main2}$(a)$
shows that two-wave soliton branch approaches $\alpha = 9.0$ from
the left, where it bifurcates from the $S_3$ asymptotic one-wave
family that has alternative phase between each single-soliton
component. However, we find that this is only one of a total of
{\bf four} symmetric two-wave solitons that come out of $S_3$.
There are 8 in total if you include the change of sign of both $u$
and $w$. The second bifurcates to the left from the same
(alternating phase) $S_3$ family and differs only in that the
first harmonic has the opposite sign. A representative of this
branch, corresponding to point $H$ in Fig.~\ref{main2}$(a)$ is
shown in Fig.~\ref{fig10}$(h)$. The two other branches exist for
$\alpha>9$ and bifurcate from the $S_3$ family where all peaks are
in phase (positive), and representatives are shown in
Fig.~\ref{fig14}$(u,v)$. With the increase of $\alpha$ (cascading
limit) these complex multi-humped solitons keep their general
structure intact, but become more localized. These two branches
are not shown in the bifurcation diagram (Fig.~\ref{main2}) but
their $P(\alpha)$ curves lie very close to each other and to the
$T_3$ curve to the right of the bifurcation point.

A similar bifurcation picture is observed at $\alpha = 9.0$ for
bifurcations from $S_5$ and $S_7$ one-wave families. However,
because of the increase of the number in possible one-wave
multi-soliton families themselves, the number of the corresponding
bifurcated two-component branches also increases. For the even
solitons considered in this work we have the following formula to
calculate the number of two-wave sub-families bifurcating from
one-wave $S_i$ family: $N_i=2^{(i+1)/2}$ (double that if we count
the change signs of $u$ and $v$). For example, there are 16
branches that bifurcate from $S_7$ branches which have $P = 84$ at
$\alpha=9$. Note that in the bifurcation diagram of
Fig.~\ref{main2}, in order to clutter, only branches directly
linked to the cascading limit two-wave family are shown. Close to
bifurcation points, the third harmonic components of the depicted
branches have neighboring humps of alternating sign and first
harmonic components have all humps of the same sign. Note that
these branches all bifurcate to the left of $\alpha=9$. For the
branches which bifurcate to the right not all third harmonic
neighboring humps alternate in sign.

\begin{figure}[h]
\vspace*{3mm} \setlength{\epsfxsize}{8cm}
\centerline{\epsfbox{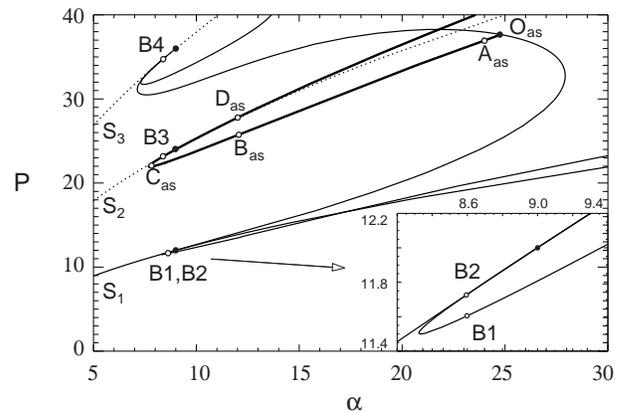}} \caption{Bifurcation diagram
from the first three one-component families $S_i, i=1,2,3$.
Asymmetric family $S_2$ is shown by a thick line.}
\label{figmain2}
\end{figure}

\begin{figure}[h]
\setlength{\epsfxsize}{7.5cm}
\centerline{\epsfbox{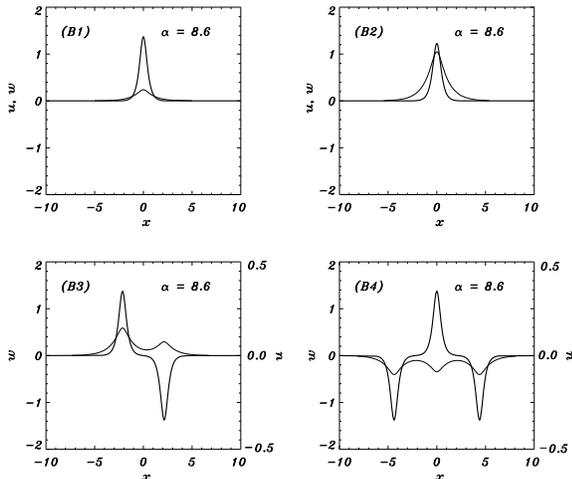}} \caption{Examples of the
two-wave solitons close to bifurcation point at $\alpha=9$. Weak
component $u(x)$ is enlarged in two bottom plots. Labeling of the
profiles is in agreement with Fig.~\ref{figmain2}.} \label{nlb}
\end{figure}

It is important to note that none of the multi-hump soliton
branches bifurcating to the left of $\alpha=9$ can be viewed as
bound states of single partial solitons. Indeed, single one-hump
solitons of Eqs.~(\ref{eq_u}) {\em always} have $u$ and $w$
components in-phase (of the same sign) for $\alpha < 9.0$, whereas
some of the individual humps of the of multi-hump structures
bifurcating to the left from $S_i$ ($i > 1$) families have $u$ and
$w$ components of different signs. To illustrate this point we
show in Fig.~\ref{figmain2} an enlarged bifurcation diagram in the
vicinity of $\alpha=9$ covering the first three families, $S_i$,
$i=1,2,3$. Some of the corresponding examples of soliton profiles
plotted at $\alpha=8.6$ are given in Fig.~\ref{nlb}. As they
approach $\alpha = 9.0$ the separation between each individual
hump (a `partial soliton') increases and the state begins to
approach a concatenation of single solitons with slightly
overlapping tails. However, some of these partial solitons have
out-of-phase $u$ and $w$ components and hence {\em cannot exist}
on their own (i.e.\ without being in superposition with other
`partial' solitons).

\begin{figure}[h]
\setlength{\epsfxsize}{7.5cm}
\centerline{\epsfbox{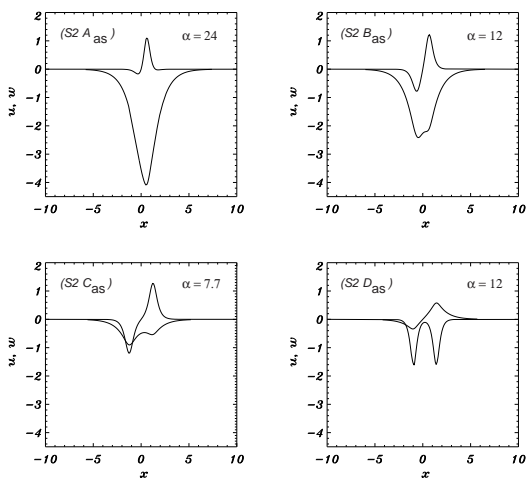}} \caption{Examples of
asymmetric solutions bifurcated from the family $S_2$. Labeling of
the profiles is in agreement with Fig.~\ref{figmain2}.} \label{t2}
\end{figure}

Fig.~\ref{figmain2} shows something even more striking - that
there is also a `bifurcation' from the $S_2$ family. However, the
solitary waves that bifurcate from there are not bright symmetric
but in fact are {\em asymmetric} solitons, see Fig.~\ref{t2}. Also
at least one of these asymmetric solutions is born in a
symmetry-breaking (pitchfork) bifurcation from one of the
symmetric soliton branches (at the point $O_{as}$, see
Fig.~\ref{figmain2}). Thus there is a branch of asymmetric
solitons which connects symmetric solitons with a branch of
asymptotic antisymetric solitons (the $S_2$ family). We conjecture
that there are similar asymmetric solitons that `bifurcate' from
$S_j$ at $\alpha=9$ for all even $j$.

In contrast to the (2+1)D case, we have found no examples (at
least considering all bifurcations from $S_{2i+1}$ with $2i+1\leq
7$) of two-wave solitons that survive down to $\alpha=0$ where
they might form a connection with branches of quasi-solitons
existing for $\alpha<0$. Instead a representative branch coming
from $T_7$ bends abruptly (at R) at which point $\alpha$ increases
through the point S until it reaches T at $\alpha \approx 3.65$,
where another nonlocal bifurcation occurs. In this process, the
third harmonic gradually forms a core with weakly separated wings.
At T ,The latter become completely separated one-wave solitons
[see Fig.~\ref{fig13}(s,t)]. The solution at the point T can thus
be viewed as a direct sum of two well-separated one-wave solitons
and the soliton at point N. Beyond $T$ we were unable to find any
similar solutions. This non-trivial ``jump'' bifurcation is
indicated by the vertical arrow in Fig.~\ref{main2}.

\section{Conclusion}

In conclusion, we have investigated and classified higher-order
soliton families and bifurcation phenomena due to resonant
parametric interaction of a fundamental frequency wave with its
third harmonic.

In the case of (2+1)D solitons the picture is consistent with
standard theories, albeit the branch we followed from the
cascading limit connects several distinct soliton types in a
non-trivial way. Also the structure of the sets of branches we
found to approach the limit $\alpha=0$ could do with further
investigation, perhaps using singular perturbation theory. The
relation of these states for positive $\alpha$ to quasi-solitons
for negative $\alpha$ will be addressed elsewhere.

In contrast, in the (1+1)D case the bifurcation diagram is less
clear-cut and we have found at least two novel features (i) the
non-local bifurcation of multi-humped two-frequency solutions
which are a consequence of the local bifurcation from the
one-humped one-frequency soliton at $\alpha=9$, and (ii) the
so-called {\em jump} bifurcation at the point $T$.  The first of
these is particularly intriguing since not only are symmetric
multi-humped states formed in this way, but also asymmetric ones.
Also some of the multi-humped states cannot be viewed as bound
states of several distinct one-humped states. The second novel
bifurcation, the jump, appears related to, but not the same as,
the so-called {\em orbit-flip} bifurcation \cite{SaJoAl:96}. A
dynamical-systems-theory explanation of these new bifurcation
events, perhaps using the Lin-Sandstede method as in
Ref.~\cite{yew}, would be most interesting.

Stability of the newly discovered soliton families remains an open
question, especially for the (1+1)D case. Although usually
higher-order soliton families are subject to one of several types
of instability, some exceptions are known (see, e.g., \cite{ostr})
and thus a careful stability analysis is worth doing. The promise
of detecting stable multi-hump solitons is {\em real} indeed
because at least some of them cannot be viewed as bound states of
two or more single (one-hump) solitons. For such bound state
solitons of NLS-type system of equations, there is practically no
hope of stability  as shown e.g. in Ref. \cite{bs}.

The authors acknowledge the use of computing facilities at Optical
Sciences Centre, RSPhysSE, the Australian National University. AVB
and RAS are indebted to O.~Bang, L.~Berge, B.~A.~Malomed, and
D.~Skryabin for useful discussions and interest in this work. ARC
is indebted to the hospitality of Yu.~Kivshar at the RSPhysSE and
to the UK EPSRC with whom he holds an advanced fellowship.

\end{multicols}
\end{document}